\documentstyle[12pt, epsfig]{article}

\def\Journal#1#2#3#4{{#1} {\bf #2}, #3 (#4)}

\def\NPB{{\em Nucl. Phys.} B}

\def\PLB{{\em Phys. Lett.}  B}

\def\PRD{{\em Phys. Rev.} D}

\begin{document}
\sloppy
\begin{titlepage}
\samepage{
\setcounter{page}{0}
\vspace{-1.in}
\rightline{Edinburgh 96/14}
\rightline{DFTT 35/96}
\rightline{hep-ph/9607291}
\vspace{.2in}
\begin{center}
{\bf DOUBLE SCALING VIOLATIONS\\}
\vspace{.3in}
{Stefano Forte\\}
\vspace{.15in}
{\it INFN, Sezione di Torino,\\ Via P.~Giuria 1, I-10125 Torino, Italy\\}
\vspace{.15in}
{Richard D. Ball\footnote{Royal Society University
Research Fellow}\\}
{\it Department of Physics and Astronomy, University of Edinburgh,\\
Edinburgh EH9 3JZ, Scotland\\}
\end{center}
\vspace{.25in}

\begin{abstract}

We discuss the theoretical implications of the
scaling properties  of $F_2^p$ at small $x$ observed in recent HERA
experiments. We show that low $Q^2$ data display double scaling 
violations which are adequately described by NL $\ln Q^2$ 
corrections. Scaling violations due to summations of leading and 
subleading $\ln {1\over x}$ beyond NLO in $\ln Q^2$ are however 
disfavoured by the data. 
 
\end{abstract}
\vspace{.2in}
\begin{center}
  Presented at the Workshop {\bf DIS 96},\\
Rome, April 1996\\
\vspace{.1in}
{To be published in the proceedings}
\end{center}
\vfill
\leftline{July 1996}
}
\end{titlepage}
\setcounter{footnote}{0}

The behaviour of the structure function $F_2^p$ as a function of $x$ and
$Q^2$ has been determined~\cite{hone}  by the H1
 collaboration to an accuracy and with a kinematic coverage
which makes precision tests of perturbative QCD now possible.
 Whereas the bulk of the data confirm the previously established
double asymptotic scaling behaviour~\cite{das}
predicted by perturbative QCD~\cite{DGPTWZ} at the leading log 
level, a study of scaling violations allows one to  determine the
perturbative mechanism which drives the evolution of $F_2$ as well
as to
disentangle it from its input nonperturbative shape:
in particular, it is now possible to establish the respective roles
of  leading logs of $Q^2$ and $1\over x$ in the QCD evolution equations
in the HERA kinematic region.

The contribution  of all logarithmic terms of the form
\begin{equation}
\alpha_s^p (\log Q^2)^q(\log{1\over x})^r.
\label{eq:logs}
\end{equation}
 to the evolution of
structure functions can be included by solution of appropriate 
renormalization group equations. 
Double asymptotic scaling 
follows from the symmetric summation of all terms with $p=q=r$. The
recent H1 data, rescaled by  this 
prediction,\footnote{The first subasymptotic correction~\cite{das}
to double scaling is also included in the rescaling factor $R_F$.}
are plotted in fig.~1 versus the product of the two scales 
$\sigma=\sqrt{\ln {t\over t_0}\ln{x_0\over x}}$, in which
$\ln F_2$ should grow linearly with a calculable universal
slope, and their ratio 
$\rho=\sqrt{\ln{x_0\over x}/\ln {t\over t_0}}$, of which
$F_2$ should be independent (with $t=\ln{Q^2\over \Lambda^2}$). 
Whereas the data agree very well
asymptotically  with double 
scaling, the observed rise of $F_2$ with $\sigma$
 appears to
be subasymptotically somewhat smaller
(i.e. the rescaled
$F_2$ drops); further scaling violations are displayed by data with lower
values of $Q^2$.
\begin{figure}[t]\begin{center}\vskip-5.truecm
\mbox{\epsfig{figure=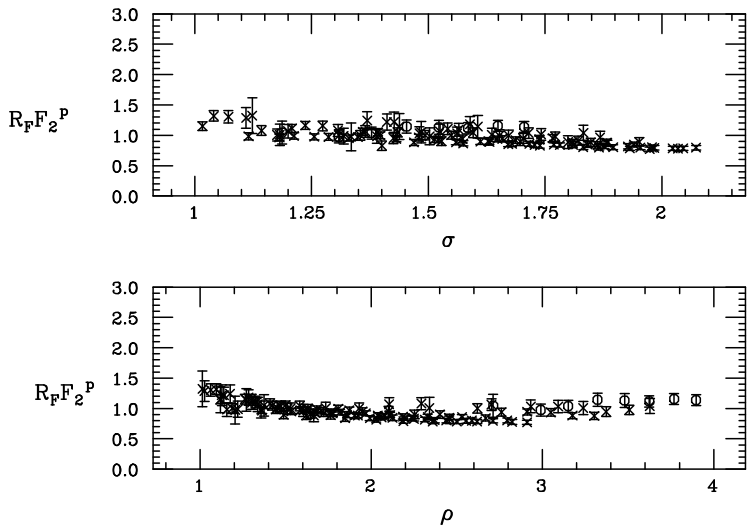,height=15.truecm}}\vskip-4.5truecm
\end{center}
\caption[]{\footnotesize
The H1 data~$^{1}$ rescaled to the LO double asymptotic
scaling prediction. The rescaling factor
is as in ref.~2. 
Only data with $\rho>1$, $\sigma>1$ and $Q^2\ge 5$~GeV$^2$ are shown.
The circles correspond to data with $Q^2=5$~GeV$^2$.\hfill
}
\end{figure}

At moderate values of $Q^2$,  logarithmic contributions
to the evolution equations 
with $p>q$ in (~\ref{eq:logs}) can be important. It
turns out that all contributions with $p=r=q+1$ vanish, hence the most important
correction (corresponding to the inclusion of the most singular
contribution in ${1\over x}$ to the NL $\ln Q^2$ terms) 
is given by summing terms with $p=q+1=r+1$. The predicted asymptotic
behaviour is still given by a universal double scaling form, but receives
now a NL (but scheme independent) enhancement, suppressed by a factor of
$\rho\alpha_s(Q^2)$~\cite{zako}. This correction thus leads to a 
rise of $F_2$ at large $\rho$, and a subasymptotic
reduction of around $10\%$ in the slope of the rise with $\sigma$ in the HERA 
region. Indeed, if the data are rescaled with this NLO prediction 
(fig.~2) the predominant 
double scaling violation apparent in fig.~1, namely the drop in $\sigma$, is
removed, and the slope of the rise of $F_2^p$ is now in excellent agreement with 
QCD~\cite{hone}. The NLO corrections 
also have the effect of raising somewhat the optimal value of the
starting scale.
\begin{figure*}[t]\begin{center}\vskip-5.truecm
\mbox{\epsfig{figure=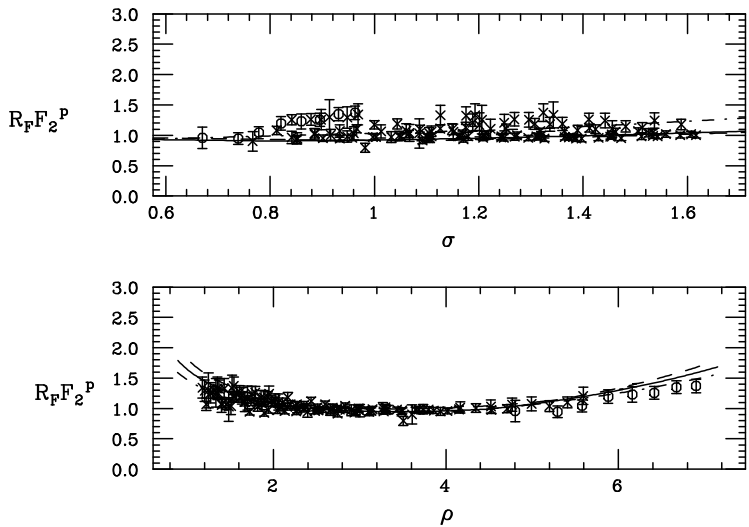,height=15.truecm}}\vskip-4.5truecm
\caption{\footnotesize
Same as fig.~1, but with NLO rescaling.
The rescaling factor is as in ref.~5, with $Q_0=1.8$~GeV. The curves
have fixed $\rho=2,3,4$ in the $\sigma$ plot and
$\sigma=$1,~1.2,~1.4 in the $\rho$ plot, and correspond to the
two loop best fit. \hfill}
\end{center}\end{figure*}

The remaining scaling violations displayed in fig.~2 could be simply
subasymptotic corrections, such as contributions to perturbative
evolution with no $\ln{1\over x}$ enhancement (i.e. $r=0$), or the influence
of the boundary conditions, or heavy quark thresholds, which can all 
be kept under control by a full two loop analysis.  If we wish to
use $F_2^p$ as measured at HERA to perform precision tests
of QCD it is important to establish whether this is
the whole story, or whether instead part of the observed 
double scaling violations may be due to higher order corrections
in $\ln {1\over x}$, since the treatment of these is subject to
sizable uncertainties related to scheme dependence~\cite{paris}.

The summation of $\ln {1\over x}$ contributions can be achieved to all
orders~\cite{summing} by
reorganising the perturbative expansion~\cite{ch},~\cite{ekl}
in such a way that $\ln {1 \over x}$ is considered
to be leading. It is thus possible to define various expansion schemes,
each of which will be more accurate in a different kinematic region.
Whereas in the usual large-$x$ 
expansion only $\ln Q^2$ is leading (i.e. in LO $p=q\ge r$, in NLO 
$p=q+1\ge r$ and so on) an expansion more tilted towards small $x$
(L-expansion) can be constructed by preserving the large
$x$ form of the LO anomalous dimensions, 
but adding to the large $x$ NLO expressions the higher order
leading singularities, i.e. terms with 
$k$ extra powers of $\alpha_s$
accompanied by $k$ powers of $\ln {1\over x}$.
It is also possible to define an expansion~\cite{summing}
(double leading or DL expansion)  that treats the two logs on the 
same footing: in LO each power of $\alpha_s$ is accompanied by either of the
two logs, while in NLO an overall extra power of $\alpha_s$ is 
allowed.\footnote{The NLO anomalous dimensions in this scheme are
known explicitly in the quark sector~\cite{ch}, but in the
gluon
sector they can be fixed  by choosing factorization
schemes in which momentum is conserved~\cite{mom}.}

Such expansion schemes will be adequate provided $x$ is small enough. They 
can then be matched to the large $x$ expansion by imposing continuity of 
anomalous  dimensions and coefficient functions at a reference value $x=x_0$.
This introduces an expansion scheme ambiguity which, once the expansion
to be used in the small $x$ region has been chosen, is parametrized by 
$x_0$.\cite{alphas} If we take for simplicity $x_0$ to be $Q^2$--independent,
this simply means that the log which is being summed is actually
$\ln {x_0\over x}$ when $x\le x_0$, so as $x_0\to0$ all schemes reduce to the 
standard loop expansion. 
It should then be possible to
determine $x_0$ by comparing the computed scaling violations to the data.
Because of the importance of leading $\ln {1\over x}$ effects already at
one and two loops in the HERA region, one might naively
expect the optimal
value of $x_0$ to be large enough that  $\ln {1\over x}$ effects
beyond two loops are already relevant there. 
Previous data did not allow~\cite{alphas} a determination of $x_0$,
essentially because even in the DL scheme the dominant asymptotic behaviour
is still given~\cite{summing} by double scaling. The recent data~\cite{hone},
~however, besides being very precise, extend well in the subasymptotic
(low $x$ and low $Q^2$) region and are thus more sensitive to $x_0$.
\begin{figure*}[t]\begin{center}\vskip-3.5truecm
\mbox{\hskip-.6truecm\epsfig{figure=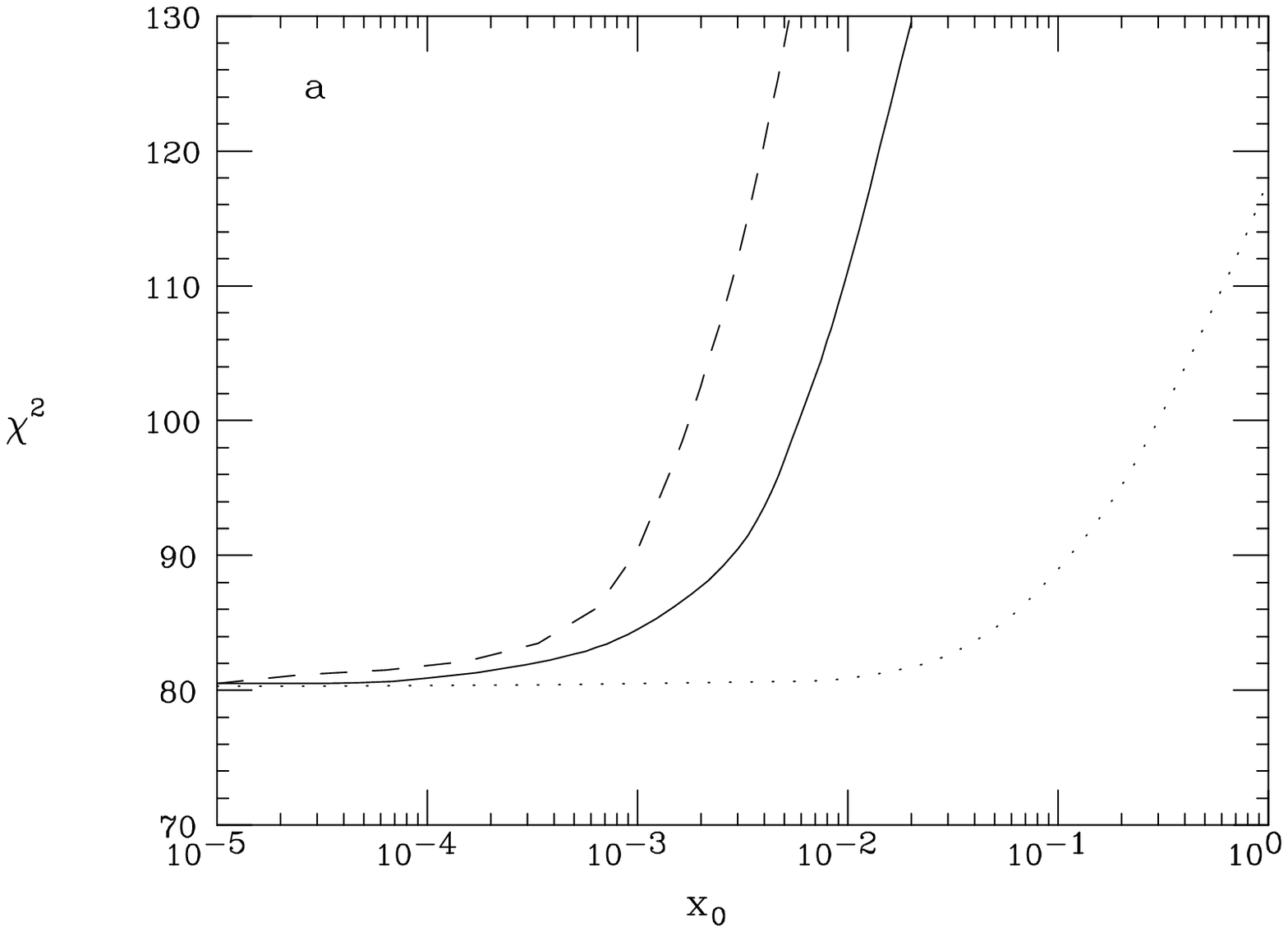,height=10.5truecm}\hskip-1.truecm
\epsfig{figure=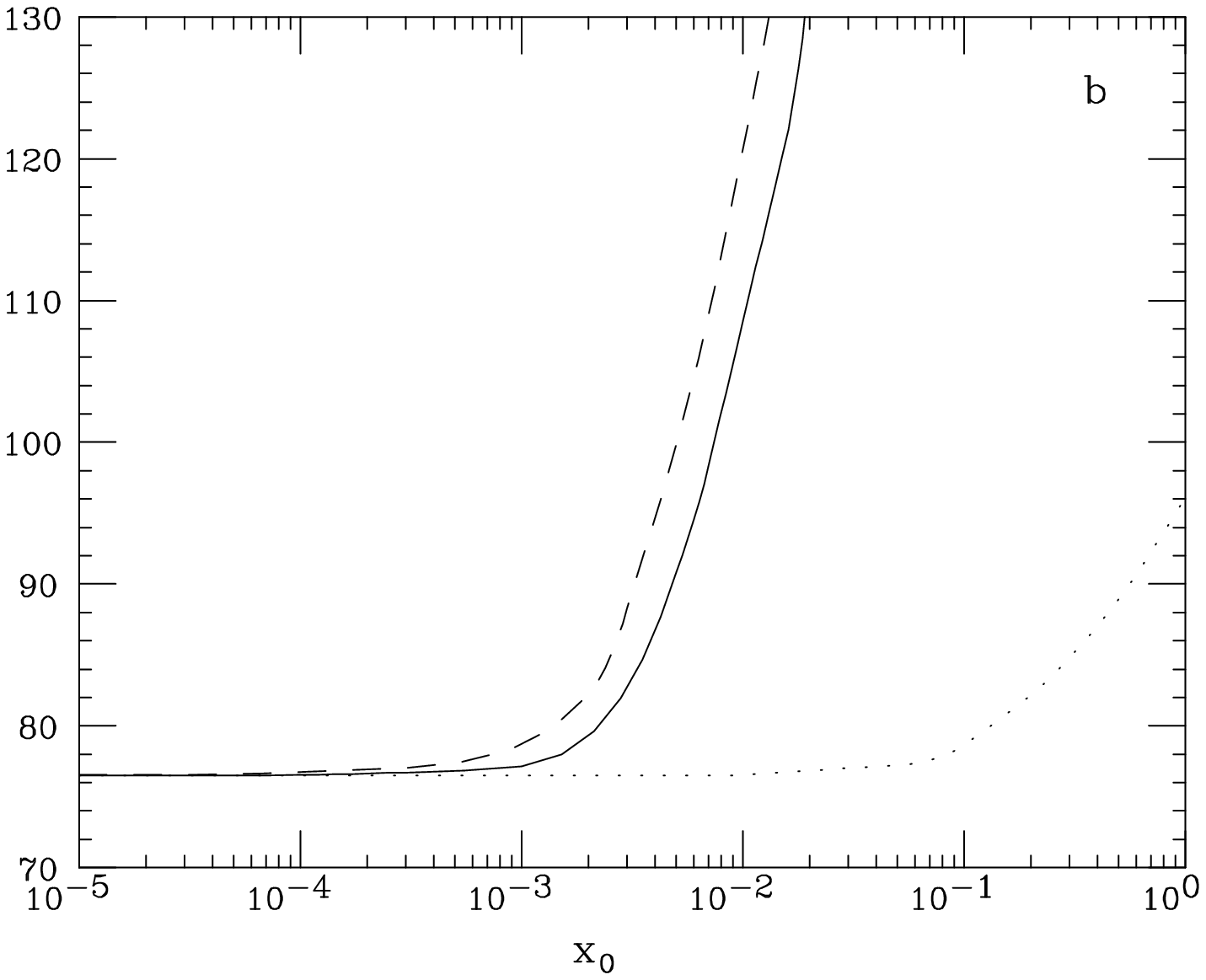,height=10.5truecm}}
\vskip-2.5truecm
\caption{\footnotesize
 The $\chi^2$ of the three parameter fit (166 d.~f.) as a 
function of $x_0$ in $\overline{\rm MS}$ (a) and DIS (b) schemes.
The solid and dashed lines correspond to the DL expansion
with standard$^{6}$ and $Q_0$$^{11}$ factorization, and
the dotted line to the L expansion.
\hfill}
\end{center}\end{figure*}

In order to search for higher order logarithms we perform fits to 
the $F_2$ data~\cite{hone} using two free parameters,
namely the 
exponents $\lambda_q$ and $\lambda_g$ which characterize
the small-$x$ behaviour $x^{\lambda_i}$ of the input singlet quark and
gluon distributions.\footnote{More details of the fitting 
procedure are given in ref.~12.}
A full set of NLO fits can then be performed with different values 
of $x_0$ and in a variety of factorization
and small $x$ expansion schemes. 

The dependence of the $\chi^2$ of these fits on the value of $x_0$ 
is displayed in fig.~3.
The values of $\lambda_q$, $\lambda_g$ and even $\alpha_s(M_Z)$ are 
refitted independently at each $x_0$.
The quality of the fit 
always gets
monotonically worse as $x_0$ is increased: 
the data generally favour a very small value 
of $x_0$. The deterioration of the fit 
is less pronounced in factorization schemes
(such as  $Q_0$~\cite{ciaf} compared to standard~\cite{ch} 
factorization)
or expansion schemes (such as L compared to DL) which are closer 
to the conventional large $x$ two loop expansion. We conclude
that the recent data~\cite{hone} seem to dislike scaling violations which go 
beyond standard two-loop effects.

We thus focus on the case $x_0=0$ (two loops). The 
results of the fit as the starting scale is varied, but with $\alpha_s$ kept fixed
at its best-fit value $\alpha_s(M_z)=0.122$~\cite{rmas}
are shown in fig.~4. The data turn out to require soft
boundary conditions, with a somewhat softer gluon distribution, 
and a moderately rising quark. As the starting scale $Q_0$ is raised, the
gluon becomes more singular and eventually leads the quark. 
The quality of the fit is largely independent
of $Q_0$ within a broad range of values. However, it deteriorates
rapidly if $Q_0$ is too high, because  the assumed power-like 
behaviour of the input spoils double scaling, or
if it is too low, because the perturbatively generated rise
becomes too strong. 
The best-fit prediction (with $Q_0=2{\rm ~GeV}$) is compared 
directly to the data in fig.2.

We can also look for higher twist contributions by repeating
the fits with $F_2$ reparametrized as  
$F_2^{HT}=F_2^{LT}(1+C_{HT}/Q^2)$. Taking $C_{HT}$ to be $x$ independent 
its best-fit value turns out to be $C_{HT}=0.2\pm 0.2$~GeV$^2$. 

In conclusion, our analysis of the 1994 H1 data suggests that
this is an ideal place to perform high-precision tests of QCD:
due to the smallness of contributions related to higher logs of $1\over x$ and
higher twists, the usual NLO perturbative expansion is perfectly adequate. 
The absence of scaling violations related to higher logs
of $1\over x$, as expressed in the 
unnaturally small value of $x_0$, is however unexpected,
suggesting that our understanding of the way these logs should be
summed is as yet incomplete.

\begin{figure*}[t]\begin{center}\vskip-3.5truecm
\mbox{\hskip-.9truecm\epsfig{figure=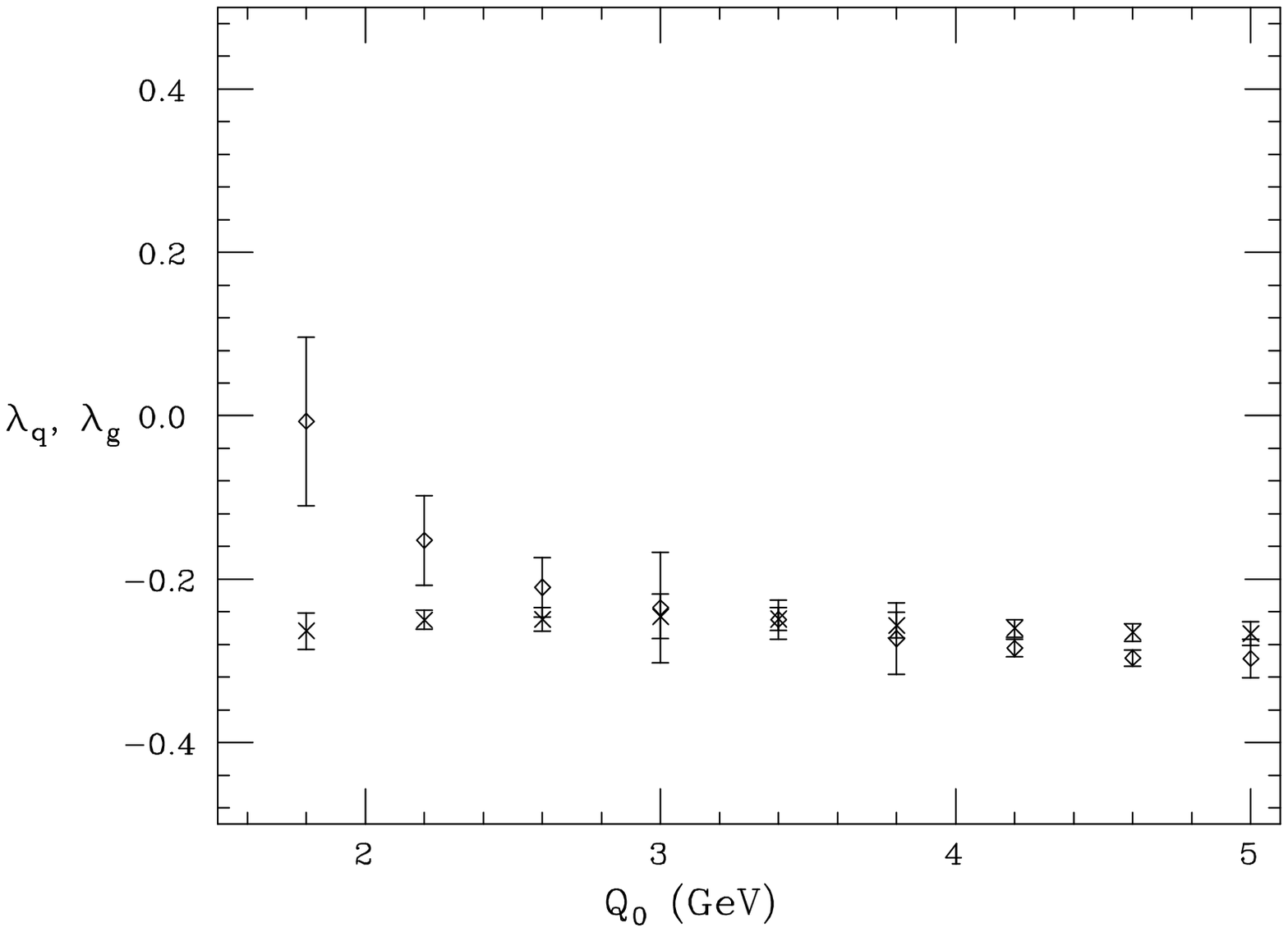,height=10.5truecm}
\hskip-1.truecm\epsfig{figure=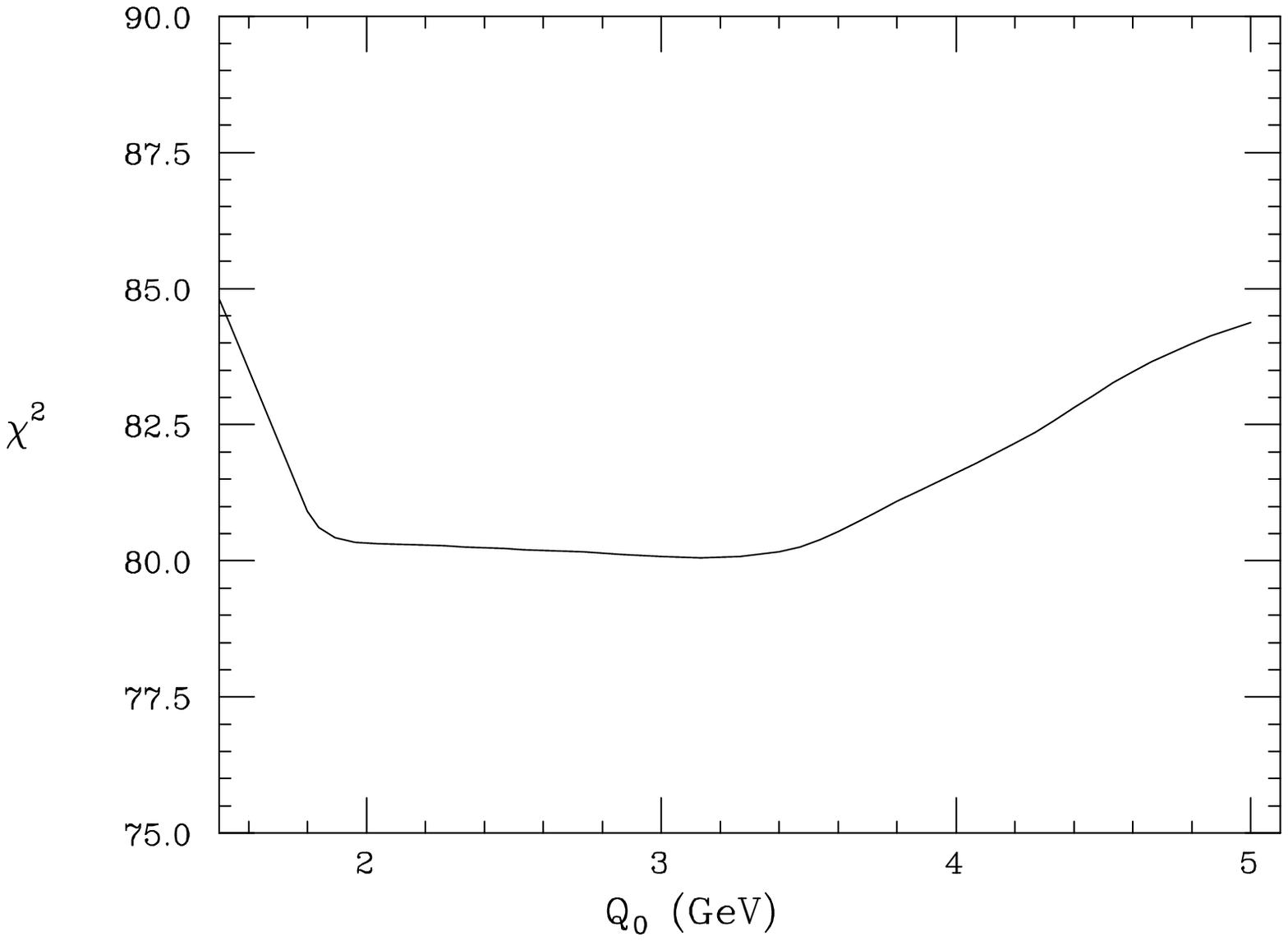,height=10.5truecm}}\vskip-3.truecm
\caption{\footnotesize
The best-fit values of the low-$x$ exponents
$\lambda_q$ (crosses) and $\lambda_g$ (diamonds) and the $\chi^2$
of the fit as a function of the starting scale $Q_0$.
\hfill}
\end{center}\end{figure*}


{
 }

\end{document}